\documentclass{emulateapj}

 \usepackage{graphics}
\usepackage{epsfig}
\usepackage{natbib}
\shorttitle{COS Observations of SNR LMC N132D}
\shortauthors{France et al.}

\begin{document}


\title{Cosmic Origins Spectrograph Observations of the Chemical Composition 
of LMC N132D.\altaffilmark{1}}


\author{Kevin France\altaffilmark{2}, Matthew Beasley, Brian A. Keeney, Charles W. Danforth, 
Cynthia S. Froning, and James C. Green }

\affil{Center for Astrophysics and Space Astronomy, 389 UCB, University of Colorado, 
Boulder, CO 80309}



    

\altaffiltext{1}{Based on observations made with the NASA/ESA $Hubble$~S$pace$~$Telescope$, obtained from the data archive at the Space Telescope Science Institute. STScI is operated by the Association of Universities for Research in Astronomy, Inc. under NASA contract NAS 5-26555.}

\altaffiltext{2}{kevin.france@colorado.edu}




\begin{abstract}
We present new far-ultraviolet spectra of an oxygen-rich knot in the Large Magellanic 
Cloud supernova remnant N132D, obtained with the $Hubble$~$Space$~$Telescope$-Cosmic Origins Spectrograph.  Moderate resolution ($\Delta$$v$~$\approx$~200 km s$^{-1}$) spectra in the 
$HST$ far-ultraviolet bandpass (1150~$\lesssim$~$\lambda$~$\lesssim$~1750~\AA) show emission from several ionization states of
oxygen as well as trace amounts of other species.  We use the improvements in sensitivity and resolving power offered by COS to separate contributions from different velocity components within the remnant, as well as emission from different species within the oxygen-rich knot itself.  This is the first time that compositional and velocity structure in the ultraviolet emission lines from N132D has been resolved, and we use this to assess the chemical composition of the remnant.  No nitrogen is detected
in N132D and multiple carbon species are found at velocities inconsistent with the 
main oxygen component.  We find helium and silicon to be associated with the oxygen-rich knot, 
and use the reddening-corrected line strengths of \ion{O}{3}], \ion{O}{4}], \ion{O}{5}, and \ion{Si}{4}
to constrain the composition and physical characteristics of this oxygen-rich knot.  We find that 
models with a silicon-to-oxygen abundance ratio of 
$N$(Si)/$N$(O)~=~10$^{-2}$ can reproduce the observed emission for a shock velocity of~$\approx$~130 km s$^{-1}$, implying a mass of $\sim$~50~$M_{\odot}$ for the N132D progenitor star.

\end{abstract}



\clearpage


\section{Introduction}

Oxygen-rich supernova remnants (SNRs) are a category of young remnants characterized by 
enhanced abundances of oxygen, neon, and other heavy elements, and are observed 
to have large outflow velocities ($v$~$\geq$~1000~km s$^{-1}$).  O-rich remnants are thought to be the ejected stellar interiors from helium-burned layers of massive ($\geq$ 10~M$_{\odot}$) progenitor stars.
These objects are ideal for studies of the evolution of massive stars and stellar
nucleosynthesis.  O-rich SNRs are found locally (e.g., Cassiopaiea A; Chevalier \& Kirshner, 1979) as well as in nearby galaxies (e.g., \citealt{blair00}).~\nocite{chevalier79}

Far-ultraviolet (far-UV) spectra of these objects provide a rich set of 
diagnostic lines for determining the composition and physical state of the stellar material. 
Far-UV observations probe excitation  
mechanisms and shock physics, providing a quantitative observational basis
for studies of the interaction between supernovae and the interstellar medium.
However, the accessibility of these objects varies widely depending on the  
foreground extinction to a given SNR.  The majority of Galactic SNRs are strongly  
reddened, making studies of local O-rich ejecta challenging.  O-rich  
SNRs beyond the Local Group of galaxies cannot be spatially resolved.  
The Magellanic Clouds offer the best compromise
of low reddening and spatially resolvable targets suitable for far-UV investigations.
Well studied O-rich remnants include N132D and SNR 0540 in the LMC, and E0102 in the  
SMC.   


The young, oxygen-rich SNR N132D is located in the bar of the Large  
Magellanic Cloud (LMC) and was first identified as a supernova remnant  
by \citet{westerlund66}.   \citet{danziger76} and \citet{lasker78}  
confirmed its oxygen-rich nature.  Spectroscopic studies (\citealt 
{danziger76,lasker80, dopita84}) revealed high-velocity filaments  
showing optical emission from oxygen and neon only spanning a total  
velocity range of $\sim$~4400 km s$^{-1}$.  The oxygen-rich filaments are  
concentrated near the middle of the remnant~\citep{borkowski07}.  Outside the fast-moving  
material is a bright x-ray shell (\citealt{long81, mathewson83}) that  
is associated with an optical emission line rim (\citealt{hughes94,  
blair94}) of radius $\sim$ 1 arcminute ($\sim$ 13 parsecs).  \cite 
{dickel95} observed  6 centimeter radio emission that  
coincides with the x-ray shell.

\cite{blair00} found that the abundances derived for the
ejecta in N132D roughly match models of a star with an initial
mass of 35 M$_{\odot}$  with the following condition: the O-rich  
mantle of the progenitor star did not mix with deeper O-burning layers.  
If there had been mixing between these layers, sulfur and silicon would have been added to the ejecta. They conclude that the ejecta is comprised of oxygen, neon,
carbon, and magnesium, which may indicate that the progenitor was a  
WO Wolf-Rayet star, perhaps with a mass as high as 85~$M_{\odot}$~\citep{woosley95}.

In this paper, we expand on previous observations of N132D. We present first 
results from the Cosmic Origins Spectrograph (COS), recently installed on the 
$Hubble$~$Space$~$Telescope$ ($HST$).  We use these data to resolve the 
composition of the O-rich knot in N132D, and evaluate the mass of the progenitor star 
based on the elemental abundances of the stellar ejecta.
We describe the COS observations and custom data reduction in \S2.  A quantitative
analysis of the far-UV spectrum is presented in \S3.  In \S4, we compare the measured 
line strengths to shock models to constrain the post-explosion abundances and 
the mass of the N132D progenitor star.

\section{$HST$-COS Observations and Data Reduction}

The Cosmic Origins Spectrograph is a slitless, modified Rowland Circle spectrograph 
designed for high-sensitivity observations of point sources in the vacuum-ultraviolet 
bandpass (1150~--~3200~\AA).  COS is optimized for deep, moderate-resolution spectroscopy
(R~$\cong$~20,000; $\Delta$$v$~$\approx$~15 km s$^{-1}$ for point source observations) in the far-UV (1150~$\lesssim$~$\lambda$~$\lesssim$~1750~\AA).  The far-UV channel employs a single optical element for dispersion, focus, and correction of the 
spherical aberration of the $HST$ primary mirror.  A pre-flight review of COS
can be found in~\citet{green03} and a full instrument description and on-orbit 
performance characteristics are in preparation (Green et al.; Osterman et al. -
in prep). ~\nocite{green10,osterman10}

\begin{figure}
\begin{center}
\hspace{+0.0in}
\epsfig{figure=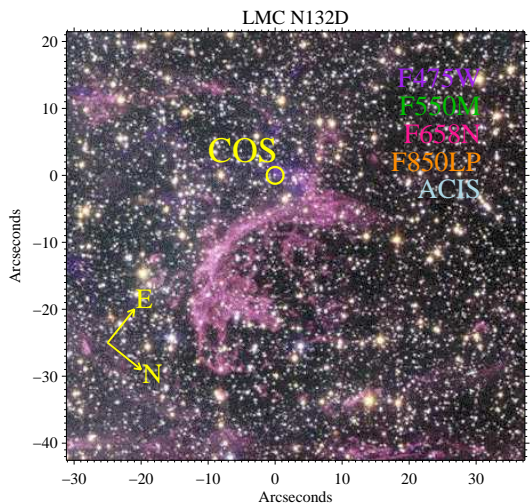,width=3.0in,angle=90}
\caption{\label{cosovly} The inner shock region of the N132D SNR observed 
by the Cosmic Origins Spectrograph.  The COS aperture is shown overlaid on the O-rich knot shown at the origin (R.A. = 05$^{\mathrm h}$ 25$^{\mathrm m}$ 25.51$^{\mathrm s}$, 
Dec. = -69\arcdeg\ 38\arcmin\ 14.0\arcsec; J2000) of this composite optical/X-ray image from 
$HST$-ACS and $Chandra$-ACIS.
 }
\end{center}
\end{figure}

COS was installed on $HST$ during STS-125/Servicing Mission 4 (SM4) in 2009 May.
The O-rich knot in N132D was observed by COS on 2009 August 10 and 31 as part of the 
Early Release Observation program ($HST$ proposal ID 11503; datasets {\tt lacc01} and {\tt lacc51}).  Spectroscopic data was obtained on N132D over 5 orbits, using both far-UV medium-resolution gratings (G130M and G160M) to provide
continuous spectral coverage from $\sim$~1150~--~1750~\AA.  The pointng (R.A. = 05$^{\mathrm h}$ 25$^{\mathrm m}$ 25.51$^{\mathrm s}$, 
Dec. = -69\arcdeg 38\arcmin 14.0\arcsec ; J2000) is coincident with 
previous far-UV spectroscopic observations made by $IUE$ (position `P1' of 
Blair et al. 1994) and $HST$-FOS (position `P3' of Blair et al. 2000).~\nocite{blair94,blair00}
The pointing is shown in Figure 1, with the COS aperture (2.5\arcsec\ diameter) overlaid on a 
composite $HST$/$Chandra$ image of the region.  The total exposure
times were 5190 and 4770 seconds for the G130M and G160M modes, respectively.  
In order to achieve continuous spectral coverage and minimize fixed pattern noise, observations
in each grating were made at four separate central wavelength settings ($\lambda$1291,
1300, 1309, and 1318 in G130M and $\lambda$1589, 1600, 1611, and 1623 in G160M).

\begin{figure}
\begin{center}
\hspace{+0.0in}
\epsfig{figure=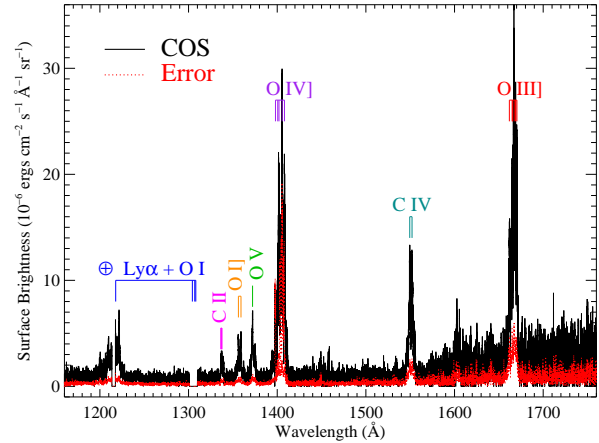,width=2.5in,angle=90}
\caption{\label{cosfull} The full COS spectrum of N132D-P3, obtained with the 
G130M and G160M modes.  Several ionization states of oxygen are observed and emission from \ion{Si}{4} (blended with \ion{O}{4}]~$\lambda$1400; Figure 3) is detected for the first time.  The oxygen and silicon are found at a common velocity; however, a velocity offset of the carbon species suggests that the carbon is spatially distinct from the O-rich knot.  Lines from geocoronal \ion{H}{1} and \ion{O}{1} have been removed.
}
\end{center}
\end{figure}

Observations made during the SM4 observatory verification period (SMOV) 
were not processed with the complete, flight-qualified version of the COS calibration
pipeline, CALCOS\footnote{We refer the reader to the COS Data Handbook for more details:~{\tt www.stsci.edu/hst/cos/documents/handbooks/
datahandbook/COS\_longdhbcover.html}}.
Custom processing was performed to properly extract, calibrate and coadd the N132D data set.
The detector coordinates used by CALCOS for extracting 1-dimensional ($\lambda$ vs. $F_{\lambda}$)
spectra from the 2-dimensional spectrogram are optimized for point sources, hence custom extraction windows were used to define proper target and background regions for this filled-aperture
observation.  A correction for spectral drift during the exposure was made by referencing an on-board calibration lamp.  Finally, pulse height screening was performed in order to remove detector hot spots and other spurious events on the microchannel plates from the final spectrum.  Spectra from individual exposures were then cross-correlated and coadded onto a common wavelength grid.  The full far-UV spectrum of N132D is shown in Figure 2 with nebular emission lines labeled.  The minimum spectral resolution is observed to be $\Delta$$v$~$\sim$~200 km s$^{-1}$, corresponding to the expected COS extended source resolving power for the medium-resolution grating modes (R~$\approx$~1500).

\section{Analysis and Results} 

\subsection{Nebular Species and Velocity Separations}
The narrow-band WFPC2~\citep{morse96} and multi-band ACS images (Figure 1) clearly indicate that the 
O-rich knot targeted by COS has a substantially different composition than other regions of the 
remnant.  This is attributable to the degree of mixing between the stellar ejecta and the ambient
pre-supernova medium, both interstellar and from earlier mass loss episodes of the
progenitor star.  The sensitivity and spectral resolving power of COS enable us to 
separate the emission from individual species unambiguously and quantify the velocity offsets between these components.  We find oxygen to dominate the emission from this knot, consistent with 
previous findings~\citep{blair94,morse96,blair00,borkowski07}.  We observe oxygen in four ionization
states (\ion{O}{1}, \ion{O}{3}, \ion{O}{4}, and \ion{O}{5}), as well as emission from \ion{C}{2}
$\lambda\lambda$1334/1335, \ion{C}{4} $\lambda\lambda$1548/1550, smaller amounts of \ion{Si}{4} $\lambda\lambda$1394/1403, \ion{He}{2} $\lambda$1640, and several unidentified lines (Table 1).

Line centers, widths, and integrated strengths are determined from Gaussian fits to the data.
We employ a modified version of the MPFIT IDL routine that accommodates interactive fitting
of multiple lines simultaneously.  On-orbit flux calibration files were not available during SMOV, and we added 25\% to the error in the line-strengths quoted in Table 1.
In order to establish the velocity of the O-rich knot, we 
assumed that the higher ionization states of oxygen were representative.  The heliocentric velocity of the \ion{O}{3}, \ion{O}{4}, and \ion{O}{5} components is 181 $\pm$ 29 km s$^{-1}$.  Within the uncertainties of the fitting procedure, this implies that the \ion{O}{1}, \ion{Si}{4}, and \ion{He}{2} are all at the velocity of knot, which we interpret as evidence for the cospatiality of these species.
A 56~\AA\ window of the spectrum including \ion{O}{5}, \ion{Si}{4}, and \ion{O}{4} is shown
in Figure 3, with dashed lines indicating the expected locations of the lines relative to 
their rest wavelengths.  

\begin{figure}
\begin{center}
\hspace{+0.0in}
\epsfig{figure=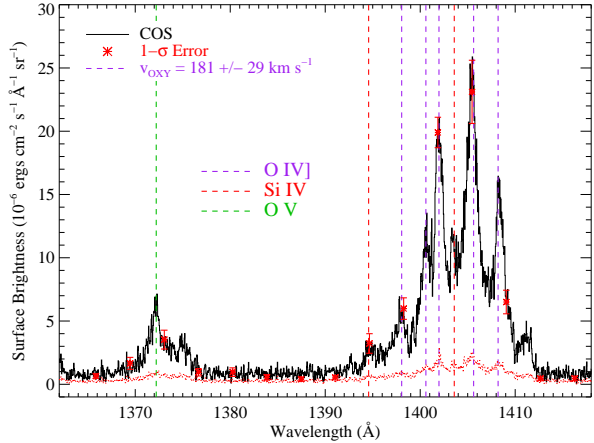,width=2.5in,angle=90}
\caption{\label{cosoiv} The spectrum of N132D in the 1362~--~1418\AA\ window.  
The data allow us to 
separate the contributions from oxygen and silicon in this knot for the first time.
The different ionization states of oxygen as well as the emission from \ion{Si}{4} 
are located at a common velocity, listed in Table 1. 
}
\end{center}
\end{figure}

The carbon components are found at significantly different velocities, 461 and 258 km s$^{-1}$
for $v_{CII}$ and $v_{CIV}$, respectively.  In addition to the separation seen in 
the carbon species, we also find weaker high-velocity emission components on the red wing of three 
line complexes.  There are high-velocity components associated with \ion{C}{4}, \ion{O}{4}, and \ion{O}{5}, all consistent with a velocity of 788 km s$^{-1}$.  Interestingly, we do not observe a 
high-velocity component in \ion{O}{3}.  
Nitrogen is not detected in any ionization state in our dataset.  We place an upper limit on the 
line flux from \ion{N}{5} $\lambda\lambda$1238/1242 of 2~$\times$~10$^{-6}$ ergs cm$^{-2}$ s$^{-1}$ sr$^{-1}$, a factor of $\sim$~30 lower than the upper limit presented by~\citet{blair00}.

We take these distinct velocity components as representative of the composition of the O-rich knot.  The knot is predominantly oxygen, with trace abundances of helium and silicon.  The carbon species and the emission associated with the high-velocity components are assumed to be spatially distinct from the O-rich knot and located in a different region of the remnant.

\begin{deluxetable}{lcccc}
\tabletypesize{\footnotesize}
\tablecaption{Observed atomic line strengths of the O-rich knot in N132D. \label{n132d_lines}}
\tablewidth{0pt}
\tablehead{
\colhead{Species\tablenotemark{a}} & \colhead{$\lambda_{obs}$} & 
\colhead{Line Flux\tablenotemark{b}}   
& \colhead{FWHM\tablenotemark{c}}  &  \colhead{$\Delta$$v$\tablenotemark{d}} \\ 
    & (\AA) &  & (km s$^{-1}$) & (km s$^{-1}$) }
\startdata
\ion{unID}{0}	& 	1221.62 	& 	3.84$\pm$0.81	 	&	 258	 & ... \\
\ion{C}{2}	& 	1336.64 	& 	1.23$\pm$0.66 		&	 172	 & 476 \\
\ion{C}{2}	& 	1337.69 	& 	1.42$\pm$0.69	 	&	 326	 & 445 \\
\ion{O}{4}]\tablenotemark{e}	& 	1339.17 	& 	1.42$\pm$0.19 		
											&	 202	 & 126 \\
\ion{O}{1}]	& 	1356.30 	& 	3.12$\pm$0.66 		&	 230	 & 155 \\
\ion{UnID}{0}	& 	1357.71 	& 	1.17$\pm$0.92 		&	 141	 & ... \\
\ion{O}{1}]	& 	1359.08 	& 	3.08$\pm$0.54 		&	 183	 & 126 \\
\ion{O}{5}	& 	1372.20 	& 	9.18$\pm$0.66 		&	 477	 & 197 \\
\ion{O}{5}$r$	& 	1374.83 	& 	3.73$\pm$0.87	 	&	 390	 & 772 \\
\ion{Si}{4}	& 	1394.86 	& 	3.01$\pm$0.87	 	&	 360  & 237 \\
\ion{O}{4}]	& 	1397.92 	& 	10.29$\pm$1.23		&	 541	 & 149 \\
\ion{O}{4}]	& 	1400.62 	& 	15.61$\pm$1.44 	&	 320	 & 245 \\
\ion{O}{4}]	& 	1402.03 	& 	18.68$\pm$1.68 	&	 210	 & 186 \\
\ion{Si}{4}	& 	1403.44 	& 	12.89$\pm$2.61 	&	 287	 & 143 \\
\ion{O}{4}]	& 	1405.45 	& 	39.86$\pm$1.32 	&	 394	 & 137 \\
\ion{O}{4}]	& 	1408.29 	& 	24.25$\pm$0.78 	&	 417	 & 194 \\
\ion{O}{4}]$r$	& 	1411.10 	& 	4.08$\pm$0.56	 	&	 281	 & 793 \\
\ion{C}{4}	& 	1549.48 	& 	16.98$\pm$2.37 	&	 377	 & 248 \\
\ion{C}{4}	& 	1552.16 	& 	19.83$\pm$2.70 	&	 476	 & 267 \\
\ion{C}{4}$r$	& 	1554.91 	& 	4.58$\pm$1.32	 	&	 187	 & 799 \\
\ion{unID}{0}	& 	1603.04 	& 	6.81$\pm$1.92	 	&	 544	 & ... \\
\ion{unID}{0}	& 	1605.87 	& 	3.28$\pm$1.80	 	&	 341	 & ... \\
\ion{He}{2}	& 	1640.31 	& 	5.73$\pm$0.72	 	&	 748	 & 143 \\
\ion{O}{3}]	& 	1661.83 	& 	6.72$\pm$0.63	 	&	 216	 & 187 \\
\ion{O}{3}]	& 	1665.15 	& 	15.03$\pm$0.99 	&	 340	 & 189 \\
\ion{O}{3}]	& 	1667.19 	& 	32.60$\pm$0.87 	&	 276	 & 188 \\
\ion{O}{3}]	& 	1668.63 	& 	12.61$\pm$1.38 	&	 222	 & 164 \\
\ion{O}{3}]	& 	1670.20 	& 	23.02$\pm$0.57 	&	 353	 & 160 \\
 \enddata
\tablenotetext{a}{Emission lines labeled as $r$ trace a high-velocity component 
(788 $\pm$ 14 km s$^{-1}$) separate from the O-rich knot (181 $\pm$ 29 km s$^{-1}$).} 
\tablenotetext{b}{10$^{-6}$ ergs cm$^{-2}$ s$^{-1}$ sr$^{-1}$}
\tablenotetext{c}{The linewidths measured from a multiple Gaussian fitting procedure 
are estimated to have a 1-$\sigma$ uncertainty of $\pm$~100 km s$^{-1}$.}
\tablenotetext{d}{ Velocity shift from the rest wavelength, \\ 
$\Delta$$v$~=~($\lambda_{obs}$~--~$\lambda_{rest}$)~$\times$~($c$/$\lambda_{rest}$). }
\tablenotetext{e}{ Tentative identification. }
\end{deluxetable}

\subsection{Reddening Correction}
In order to make a meaningful comparison to the ultraviolet emission lines predicted 
by supernova remnant shock models (described in the following section), a correction must be made
for the effects of differential extinction by interstellar dust.  We carried out
an extinction correction by assuming that all of the relevant dust was associated with the 
LMC and can be represented by the average 30 Doradus reddening curve~\citep{fitz86}.  Taking
$E(B-V)$~=~0.12 with a selective extinction R$_{V}$~=~3.2~\citep{blair00}, a continuous 
extinction curve is created for the COS bandpass.  The line strengths are corrected by dividing
by this curve in transmission space.  

\section{Discussion}

\subsection{O-rich Shock Models and Physical Conditions in the Knot}

We have compared the emission line strengths observed by COS with a grid of supernova shock
models in order to constrain important physical quantities including the density of the preshock medium ($n$), the abundances of carbon and silicon relative to oxygen, and the shock velocity.
We have used a modified version of the~\citet{raymond79} shock model code.  This is the same code used to model the $HST$-FOS observations presented by~\citet{blair00}, and we refer the reader to that paper for details of the model.  We ran this model in the weakly magnetized limit, with a pre-shock 
density of $n$~=~1~cm$^{-3}$ normalized at a shock velocity of~$V_{SHOCK}$~=~100 km s$^{-1}$ for a constant ram pressure and a power-law index of $\alpha$~=~0.4.
We used the elemental abundances from~\citet{blair00} for silicon ($N$(O)/$N$(Si)~=~10$^{2}$) and from our
COS-derived upper limit for carbon ($N$(O)/$N$(C)~$\geq$~10$^{2}$).  We ran a grid of models for values 
of $v_{SHOCK}$~=~30$~-~$200~km s$^{-1}$, and tabulated the predicted far-UV emission line 
strengths in three ionization states of oxygen (\ion{O}{3}, \ion{O}{4}, and \ion{O}{5}) and
\ion{Si}{4}.  

\begin{figure}
\begin{center}
\hspace{+0.0in}
\epsfig{figure=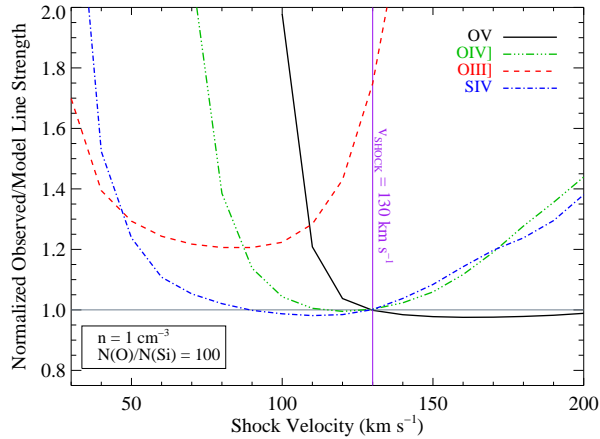,width=2.5in,angle=90}
\caption{\label{cosmod} A comparison of the observed \ion{O}{5}~$\lambda$1372, 
\ion{O}{4}]+\ion{Si}{4}~$\lambda$1400, and \ion{O}{3}~$\lambda$1664 
line strengths with those predicted by the modified Raymond et al. (1979) model
described in \S4.  The Observed/Model ratios were normalized to unity 
for \ion{O}{4} at $v_{SHOCK}$~=~130 km s$^{-1}$.  
The model confirms the pre-shock density of $n$~=~1~cm$^{-3}$ and
oxygen-to-silicon abundance ratio ($N$(O)/$N$(Si) = 100) used in~\citet{blair00}, 
and constrains the shock velocity in the highly ionized species.
We use the $N$(O)/$N$(Si) ratio in conjunction with an upper limit on 
$N$(O)/$N$(C) to estimate the N132D progenitor mass to be $\sim$~50 $M_{\odot}$.
 }
\end{center}
\end{figure}

The normalized ratios of observed/model line strengths are shown in Figure~4.  We find relatively good agreement between the model and observed line strengths with a shock velocity of $v_{SHOCK}$~=~130~$\pm$~20 km s$^{-1}$ for \ion{O}{5}, \ion{O}{4}, and \ion{Si}{4}.  The fit for the \ion{O}{3} line strength is discrepant in both shock velocity and surface brightness.  
It is interesting to note that the $\sim$~20\% offset for the \ion{O}{3} line strength is exactly the same difference in the dust attenuation curve between 1400 and 1664~\AA.  We suggest that the extinction curve to N132D is significantly flatter than the typical 30 Doradus or LMC extinction curves, 
perhaps reflecting a depleted population of small dust grains (R$_{V}$~$>$~4.0) on the sightline to this remnant.  The \ion{O}{3} appears to be more sensitive to slower shocks 
($v_{SHOCK}$~$\sim$~80 km s$^{-1}$, while the higher ionization traced by \ion{O}{5} cannot be produced in shocks slower than
$\sim$~100 km s$^{-1}$.  Taken as a whole, these findings agree with a physical scenario where shocks of different velocities are seen superposed~\citep{vancura92} in this O-rich knot, which has a preshock density and silicon abundance consistent with those presented in~\citet{blair00}, while we rule out a substantial carbon abundance in this region.

\subsection{Mass of the Progenitor Star}

The relative abundances of the O-rich knot in N132D can be compared to models of massive stars and supernova explosions to constrain the mass of the progenitor star.  It should be stressed that 
we are presenting an analysis of one small knot of ejecta in a much larger remnant.  While we can use this information to place bounds on the progenitor star, it would be inappropriate to make definitive statements regarding the nature of the presupernova star based on such a small sample of the stellar core.  With that caveat in mind, we find relatively good agreement between our abundances and those of the 40 $M_{\odot}$ main sequence model presented in~\citet{nomoto97}.  Similar results are
seen for the 70 $M_{\odot}$ star, but the lower $N$(Si)/$N$(O) of the former model more closely reproduces our results.  The 25 $M_{\odot}$ can be ruled out as it predicts too little silicon and $N$(C)/$N$(Si) greater than unity.  \citet{blair00} also comment on the 85 $M_{\odot}$ Wolf-Rayet B model~\citep{woosley93} with core masses from 2.7~$\leq$~$M_{CORE}$~$\leq$~7 $M_{\odot}$ being a reasonable fit to the observations.  This model has a relatively high $N$(C)/$N$(O) ($>$ 0.1).  In light of the spatial separation of the carbon components, which could not be resolved in the FOS data, we consider this model to be ruled out.  We note however, that for a narrow range of core mass ($M_{CORE}$~2.6~$\pm$~0.1 $M_{\odot}$), the 60 $M_{\odot}$ Wolf-Rayet A model (the distinction between ``A'' and ``B'' in Woosley et al. 1993 is the reaction rate for the conversion of carbon to oxygen in the stellar core) provides a reasonable fit to the abundances.  In this model, the relative silicon and carbon abundances are of order 10$^{-2}$, in agreement with the COS data.  Additionally, more recent models by~\citet{umeda08} calculate abundances of oxygen and silicon in the post-explosion material for 50 and 100 $M_{\odot}$ main sequence stars.  Their 50 $M_{\odot}$ model, with an explosion energy of 3~$\times$~10$^{52}$ ergs matches the $N$(Si)/$N$(O) ratio for the O-rich N132D knot, while the post-explosion yields for a 100 $M_{\odot}$ star overpredict the silicon abundance by an order of magnitude.  Given the findings described above, we adopt a mass of 50 $\pm$~25 $M_{\odot}$ for the progenitor to the N132D supernova.


\acknowledgments
We acknowledge the heroic efforts of the STS-125 crew during Servicing Mission 4 to the $Hubble$~$Space$~$Telescope$.  It is a pleasure to acknowledge Jon Morse for creating the initial observing plan for these observations and assistance with the shock modeling. KF thanks Josh Destree for data analysis assistance, and Bill Blair for insightful discussion.  We thank Eric Burgh for a careful reading of the manuscript.  This work was support by NASA grants NNX08AC146 and NAS5-98043 to the University of Colorado at Boulder.


\bibliography{n132d_em}

\end{document}